\documentclass[11pt]{article}
\usepackage{amscd,amsmath,amssymb,latexsym,bbm,graphicx}

\catcode`@=11
\makeatletter\renewcommand{\section}{\@startsection
{section}{1}{\z@}{-3.5ex plus -1ex minus
    -.2ex}{2.3ex plus .2ex}{\large\bf }}

\makeatletter\renewcommand{\subsection}{\@startsection{subsection}{2}{\z@}{-3.25ex
plus -1ex minus
   -.2ex}{1.5ex plus .2ex}{\bf }}

\numberwithin{equation}{section}
\newcounter{saveeqn}

\catcode`@=12
\def\bea{\begin{eqnarray}}
\def\eea{\end{eqnarray}}

\def\Ft{\widetilde{\cal F}}
\def\a{\alpha}
\def\b{\beta}
\def\g{\gamma}

\def\ve{\varepsilon}
\def\vk{\varkappa}
\def\r{\rho}

\def\vt{\vartheta}
\def\la{\lambda}

\def\om{\omega}
\def\1{{\bar 1}}
\def\2{{\bar 2}}
\def\3{{\bar 3}}
\def\4{{\bar 4}}
\def\ab{\bar\alpha}
\def\bb{\bar\beta}
\newcommand{\ah}{\hat{a}}
\newcommand{\bh}{\hat{b}}
\newcommand{\ch}{\hat{c}}
\newcommand{\C}{\mathbb C}
\newcommand{\R}{\mathbb R}

\newcommand{\Z}{\mathbb Z}
\newcommand{\unity}{\mathbbm{1}}
\newcommand{\U}{{\cal U}}
\newcommand{\Tcal}{{\cal T}}
\newcommand{\Acal}{{\cal A}}
\newcommand{\Ecal}{{\cal E}}

\newcommand{\Fcal}{{\cal F}}

\newcommand{\Zcal}{{\cal Z}}
\newcommand{\Jcal}{{\cal J}}

\newcommand{\hra}{\mathop{\hookrightarrow}}

\def\im{\textrm{i}}

\def\diff{\textrm{d}}
\def\tr{\textrm{tr}}
\def\sfrac#1#2{{\textstyle\frac{#1}{#2}}}
\def\+{\dagger}
\def\={\ =\ }
\def\sp{\phantom{-}}
\def\und{\qquad\textrm{and}\qquad}
\def\and{\quad\textrm{and}\quad}
\def\with{\quad\textrm{with}\quad}
\def\for{\quad\textrm{for}\quad}

\begin{document}

\begin{titlepage}
\setcounter{page}{0}
\begin{flushright}
ITP--UH--12/12\\
\end{flushright}


\begin{center}

{\Large\bf Instantons on the six-sphere and twistors }

\vspace{12mm}

{\large
Olaf~Lechtenfeld${}^{\+\times}$ \ and \ Alexander~D.~Popov${}^{*}$}
\\[8mm]
\noindent ${}^\dagger${\em
Institut f\"ur Theoretische Physik and Riemann Center for Geometry and Physics \\
Leibniz Universit\"at Hannover \\
Appelstra\ss{}e 2, 30167 Hannover, Germany }\\
{Email: lechtenf@itp.uni-hannover.de}
\\[8mm]
${}^\times${\em
Centre for Quantum Engineering and Space-Time Research \\
Leibniz Universit\"at Hannover \\
Welfengarten 1, 30167 Hannover, Germany }
\\[8mm]
\noindent ${}^*${\em
Bogoliubov Laboratory of Theoretical Physics, JINR\\
141980 Dubna, Moscow Region, Russia}\\
{Email: popov@theor.jinr.ru}

\vspace{12mm}

\begin{abstract}
\noindent We consider the six-sphere $S^6=G_2/$SU(3) and its twistor space $\Zcal =
G_2/$U(2)  associated with the SU(3)-structure on $S^6$. It is shown that a Hermitian
Yang-Mills connection (instanton) on a smooth vector bundle over $S^6$ is equivalent to
a flat partial connection on a vector bundle over the twistor space $\Zcal$. The relation
with Tian's tangent instantons on $\R^7$ and their twistor description are briefly discussed.
\end{abstract}

\end{center}
\end{titlepage}

\section{Introduction and summary}

The twistor description of solutions to chiral zero-rest-mass field equations on the
six-dimensional space~$\C^6$ or its real forms with various signatures (see e.g.~\cite{Hu})
was generalized recently to Abelian~\cite{SW1, MRT} and non-Abelian~\cite{SW2} holomorphic
principal 2-bundles over the twistor space $Q'_6\subset\C P^7\setminus \C P^3$,
corresponding to solutions of the 3-form self-duality equations on~$\C^6$.
The twistor approach was also extended to maximally supersymmetric Yang-Mills theory
on~$\C^6$~\cite{SWW}. Twistor methods have also been applied in the study of scattering
amplitudes in this theory (see e.g.~\cite{Cheung}).

The goal of our paper is to describe instantons in gauge theory on Euclidean
six-dimensional space (i.e.~the bosonic sector of maximally supersymmetric Yang-Mills theory)
by using twistor methods. Recall that instantons in four dimensions are nonperturbative
gauge-field configurations solving conformally invariant first-order anti-self-duality
equations which imply the full Yang-Mills equations~\cite{BPST}. The twistor approach allows one to
describe instanton solutions and their moduli space very efficiently~\cite{Ward, AHS, ADHM}.
We will apply it to study gauge instantons on the six-dimensional sphere~$S^6$
which is a natural compactification of~$\R^6$. Our considerations are based on papers studying
twistor spaces associated with higher-dimensional manifolds~\cite{BO}-\cite{Wo12} as well as on
papers considering instanton equations in dimensions higher than four~\cite{CDFN}-\cite{HN}.
Here, we consider instanton equations only on~$S^6$. However, our results can be
generalized to any nearly K\"ahler manifold in six and higher dimensions as well as to
some other manifolds with $G$-structure.

We recall some definitions to clarify our purposes. Let $X$ be a Riemannian
manifold of dimension~$2n$. We define the {\it metric\/} twistor space of~$X$ as the bundle
Tw$(X)\to X$ of almost Hermitian structures on~$X$ (i.e.~almost complex structures
compatible with the metric $g$ on~$X$ and its orientation) associated with the principal bundle
$P(X, \mbox{SO}(2n))$ of orthonormal frames of~$X$, i.e.
\begin{equation}
\mbox{Tw}(X):= P(X, \mbox{SO}(2n))\times_{{\rm SO}(2n)}^{}\mbox{SO}(2n)/\mbox{U}(n)\ .
\end{equation}
It is well known that Tw$(X)$ can be endowed with an almost complex structure~$\Jcal$
which is integrable if and only if the Weyl tensor of~$X$ vanishes identically when
$n>2$~\cite{BO}.\footnote{
In the case $n=2$ the Weyl tensor has to be anti-self-dual~\cite{Penrose, AHS}.}
However, if the manifold~$X$ has a $G$-structure (which is not necessarily integrable)
then one can often find a subbundle~$\Zcal$ of Tw$(X)$ associated
with the $G$-structure bundle $P(X,G)$ for $G\subset\ $SO($2n$), such that an induced
almost complex structure (also called~$\Jcal$) on~$\Zcal$ is integrable. Many examples
were considered in the literature~\cite{BO}-\cite{Sal}, \cite{AG}-\cite{BSV}.

The six-sphere~$S^6$ provides an interesting example. Considered as the round sphere
\begin{equation}
S^6\= \mbox{Spin(7)}/\mbox{Spin(6)}\ ,
\end{equation}
its metric twistor space is
\begin{equation}
\mbox{Tw}(S^6)\=\mbox{Spin(7)}/\mbox{U(3)}\ \buildrel{\C P^3}\over{\longrightarrow}\ S^6
\end{equation}
(see e.g.~\cite{Mu, BSV}), which may be recognized as a six-dimensional quadric\footnote{
Instead of~$S^6$, one can also consider $\R^6=S^6\setminus \{\infty\}$ with a metric
twistor space $Q_6'=\ $Tw$(\R^6)\subset Q_6$ as an intersection of the quadric~$Q_6$ with
$\C P^7\setminus\C P^3$. This space~$Q_6'$ is also the twistor space for the complexified
space-time~$\C^6$ with a double fibration establishing the correspondence between
subspaces in $Q_6'$ and~$\C^6$ (see e.g.~\cite{SW1, MRT} and references therein). There
is no such correspondence in the real framework where the twistor space is the bundle of
almost complex structures on a manifold.} $Q_6$ in~$\C P^7$. Alternatively, one may
consider the six-sphere as a nearly K\"ahler homogeneous space with SU(3)-structure,
namely
\begin{equation}
S^6\=G_2/\mbox{SU(3)}\ .
\end{equation}
Then,
\begin{equation}\label{1.5}
\Zcal\=G_2/\mbox{U(2)}\ \buildrel{\C P^2}\over\longrightarrow\ S^6
\end{equation}
is a complex subbundle of Tw$(S^6)$~\cite{BO, Bryant1, Sal, But}. Note that $\Zcal$ can
be identified with a five-dimensional quadric $Q_5\subset \C P^6$, and obviously
$Q_5\subset Q_6=\ $Tw$(S^6)$. The twistor space (\ref{1.5}) is a bundle of almost complex
structures $J$ on $S^6$, which are parametrized by the complex projective space $\C P^2$
at each point of $S^6$.

There was an attempt \cite{Manin}, not quite successful, to obtain instanton-type
configurations on $S^6$ from holomorphic bundles over Tw$(S^6)$. However, natural
instanton equations on $S^6$ (as well as on $\R^6$) are the Donaldson-Uhlenbeck-Yau (DUY)
equations\footnote{
In the mathematical literature they are often called Hermitian Yang-Mills equations.}
\cite{DDUY}, which are SU(3)~invariant but not invariant under the SO(6)
transformations on the round six-sphere.

The DUY~equations are well defined on $S^6=G_2/$SU(3), and their solutions are natural
connections $\Acal$ on pseudo-holomorphic vector bundles $E\to S^6$~\cite{Bryant3}.
We will show that such bundles $(E, \Acal )$ are pulled back to complex vector bundles
$(\widetilde E, \widetilde\Acal )$ over the complex twistor space $\Zcal=G_2/$U(2) with flat partial
connection\footnote{
For the definition and discussion of such connections see e.g.~\cite{Raw, PSW, Wo12}.}
$\widetilde\Acal$. The bundle $\widetilde E\to \Zcal$ is not holomorphic. We would like to emphasize
two outcomes of our study of instantons on $S^6$:\\
(i) the reduced twistor space $\Zcal\hra $Tw($X$) of~$X$ may be more suitable for
describing solutions of field equations on manifolds~$X$ with $G$-structure than the metric twistor space Tw($X$),\\
(ii) the twistor description of gauge instantons in dimensions higher than four may lead to
non-holomorphic bundles over the reduced twistor space~$\Zcal$ even if $\Zcal$ is a complex
manifold.

\section{Nearly K\"ahler structure on $S^6$}

\noindent
{\bf Almost complex structure. \ }
Let us consider the principal fibre bundle
\begin{equation}\label{2.1}
 G_2\ \longrightarrow\ G_2/\mbox{SU(3)}\=S^6
\end{equation}
with the Lie group SU(3) as the structure group.
Let $\{e^a\}$ with $a=1,\ldots,6$ be a (local) coframe on $S^6$ compatible with the SU(3)-structure
and $\{e^i\}$ with $i=7,\ldots,14$ be the
components of an $su(3)$-valued connection on the bundle (\ref{2.1}). Using $e^a$, one can introduce an
almost complex structure $J$ on $S^6$ such that
\begin{equation}\label{2.2}
 J\,\theta^\a = \im \,\theta^\a ,\ \a=1,2,3\ ,\for \theta^1:=e^1+\im\,e^2,\ \theta^2:=e^3+\im\,e^4,\ \theta^3:=e^5+\im\,e^6
\end{equation}
as well as define forms
\begin{equation}\label{2.3}
\om :=\sfrac{\im}{2}\bigl(\theta^1\wedge\theta^{\1}+\theta^2\wedge\theta^{\2}+
\theta^3\wedge\theta^{\3}\bigr)\und \Omega:=\theta^1\wedge\theta^2\wedge\theta^3\ .
\end{equation}

\bigskip

\noindent
{\bf Flat connection on $S^6$. \ }
It is convenient to work with the matrices
\begin{equation}\label{2.4}
  \theta :=\vk\,\bigl(\theta^1\,\theta^2\,\theta^{3}\bigr)\ ,\quad
\bar\theta :=\vk\,\bigl(\theta^{\1}\,\theta^{\2}\,\theta^{\3}\bigr)\ ,
\end{equation}
\begin{equation}\label{2.5}
B:=\sigma\, \begin{pmatrix}
\ 0 & \sp\theta^3 & -\theta^2 \\
-\theta^3 & \ 0 & \sp\theta^1 \\
\sp\theta^2 & -\theta^1 & \ 0 \end{pmatrix}
\und
\bar B:=\sigma\, \begin{pmatrix}
\ 0 & \sp\theta^{\3} & -\theta^{\2} \\
-\theta^{\3} & \ 0 & \sp\theta^{\1} \\
\sp\theta^{\2} & -\theta^{\1} & \ 0 \end{pmatrix}
\end{equation}
with $\vk=\sqrt{\sfrac{2}{3}}$ and $\sigma =\sqrt{\sfrac{1}{3}}$.
Using (\ref{2.4}) and~(\ref{2.5}), one can introduce a
flat Lie\,$G_2$-valued connection $\Acal_0$~\cite{Macfarline} on the trivial bundle $G_2\times S^6\to S^6$ as
\begin{equation}\label{2.6}
\Acal_0= \begin{pmatrix}\bar\Gamma&-\theta^\+&B\\\theta&\ 0&\bar\theta\\\bar B&
-\bar\theta^\+ &\Gamma\end{pmatrix} \quad\with \Gamma=e^iI_i \and \bar\Gamma=e^i\bar{I}_i= - e^iI_i^{\top}\ ,
\end{equation}
where $I_i=-I_i^\+$ are 3$\times$3 matrix generators of the group SU(3) and $\Gamma=e^iI_i$ is the canonical
connection in the bundle~(\ref{2.1}).

\bigskip

\noindent {\bf Maurer-Cartan equations on $S^6$. \ }
The flatness of the connection~(\ref{2.6}) (see~\cite{Macfarline}) means that
there exists a local $G_2$-valued function $L$ which is a coset representative of $G_2/$SU(3) such that
$\Acal_0=L^{-1}\diff\,L$. Note that $L$ is a local section of the bundle~(\ref{2.1}).
For the curvature $\Fcal_0=\diff\Acal_0 + \Acal_0\wedge\Acal_0$ we find
\begin{equation}\label{2.7}
\Fcal_0\=\begin{pmatrix}\bar{R}-\theta^\+{\wedge}\theta + B{\wedge}\bar{B}&
-(\diff\theta^\+{+}\bar\Gamma{\wedge}\theta^\+{+}B{\wedge}\bar\theta^\+)&
\diff B{+}\bar\Gamma{\wedge}B{-}\theta^\+{\wedge}\bar\theta{+}B{\wedge}\Gamma\\
\diff\theta + \theta{\wedge}\bar\Gamma + \bar\theta{\wedge}\bar B&
-(\theta{\wedge}\theta^\+ + \bar\theta{\wedge}\bar\theta^\+)   &
\diff\bar\theta +\bar\theta{\wedge}\Gamma + \theta{\wedge} B\\
\diff\bar B{+}\bar B{\wedge}\bar\Gamma{-}\bar\theta^\+{\wedge}\theta{+}\Gamma{\wedge}\bar
B& -(\diff\bar\theta^\+{+}\Gamma{\wedge}\bar\theta^\+{+}\bar B{\wedge}\theta^\+)&
{R}-\bar\theta^\+{\wedge}\bar\theta + \bar B{\wedge}{B}\end{pmatrix} \ .
\end{equation}

{}From $\Fcal_0=0$ it follows that
\begin{equation}\label{2.8}
\diff\begin{pmatrix}\theta^1\\\theta^2\\\theta^3\end{pmatrix}
+\Gamma\wedge\begin{pmatrix}\theta^1\\\theta^2\\\theta^3\end{pmatrix}\=
\frac{2}{\sqrt{3}}\begin{pmatrix}\theta^{\2}\wedge\theta^{\3}\\\theta^{\3}\wedge
\theta^{\1}\\\theta^{\1}\wedge\theta^{\2}\end{pmatrix}
\qquad\Rightarrow\qquad \diff\theta^\a +\Gamma^\a_\b\wedge\theta^\b\=T^\a\ ,
\end{equation}
where $\Gamma =(\Gamma^\a_\b)=(\Gamma^iI^\a_{i\b})$ is the canonical connection on the
tangent bundle $TS^6$ associated to the bundle~(\ref{2.1}), and where $T^\a=\sfrac12 \,
T^\a_{\bb\bar\g}\theta^{\bb}\wedge\theta^{\bar\g}$ is the intrinsic torsion of $\Gamma$
(see e.g.~\cite{Salamon}). Equation~(\ref{2.8}) and its complex conjugate constitute the
Maurer-Cartan equations on the sphere $S^6$.

The curvature $R=\diff\Gamma + \Gamma\wedge\Gamma$ of the connection $\Gamma$ is read
off~(\ref{2.7}) by equating to zero its lower right (or upper left) block,
\begin{equation}\label{2.9}
R\=\bar\theta^\+\wedge\bar\theta - \bar B\wedge
B\=\frac{1}{3}\begin{pmatrix}2\theta^{1\1}{-}\theta^{2\2}{-}\theta^{3\3}&3\theta^{1\2}&3\theta^{1\3}\\
3\theta^{2\1}&-\theta^{1\1}{+}2\theta^{2\2}{-}\theta^{3\3}&3\theta^{2\3}\\
3\theta^{3\1}&3\theta^{3\2}&-\theta^{1\1}{-}\theta^{2\2}{+}2\theta^{3\3}\end{pmatrix}\ ,
\end{equation}
where $\theta^{1\1}=\theta^1\wedge\theta^{\1}$ etc. Also, from (\ref{2.8}) we see that the
almost complex structure (\ref{2.2}) on $S^6$ is not integrable due to the torsion
$T^\a$ which is a (0,2)-form w.r.t.\ $J$. It is easy to show that
\begin{equation}\label{2.10}
\diff\om \= 3\rho\, \mbox{Im}\Omega\und \diff\Omega \= 2\rho\,\om\wedge\om \ ,
\end{equation}
where $\rho\in \R$ is proportional to the inverse radius of $S^6$. The pair $(\om
,\Omega)$ of forms subject to~(\ref{2.10}) turns $S^6$ into a nearly K\"ahler manifold
(see e.g.~\cite{But, Bryant3, Salamon}). It comes with a non-integrable SU(3)-structure.

\bigskip

\noindent {\bf Hermitian Yang-Mills equations. \ }
Consider an oriented $2n$-dimensional
Riemannian manifold~$X^{2n}$ with an almost complex structure~$J$ and a complex vector
bundle~$E$ over~$X^{2n}$ with a connection~$\Acal$. According to Bryant~\cite{Bryant3},
a connection $\Acal$ on~$E$ defines a pseudo-holomorphic
structure if it has curvature $\Fcal =\diff\Acal + \Acal\wedge\Acal$ of type (1,1)
w.r.t.\ $J$, i.e.~if $\Fcal^{0,2}=0=\Fcal^{2,0}$.

One can endow the bundle~$E$ with a Hermitian metric and choose $\Acal$ to be
compatible with the Hermitian structure on~$E$. If, in addition, $\om$ is an almost
Hermitian structure on $(X^{2n}, J)$ and $c_1(E)=0$,\footnote{
{}From a bundle with curvature $\Fcal$ of non-zero degree we can obtain a zero-degree bundle~$E$
by considering $\Ft = \Fcal - \sfrac{1}{k}\,(\tr\Fcal)\cdot{\unity}_k$, where $k=\,$rank$E$.}
then the equations
\begin{equation}\label{2.11}
\Fcal^{0,2}=-(\Fcal^{2,0})^\+=0\und \om\lrcorner\,\Fcal:=\om^{ab}\Fcal_{ab}=0
\end{equation}
are called the {\it Hermitian Yang-Mills equations\/}. The notation $\om\lrcorner\,$
exploits the underlying Riemannian metric $g=\delta_{ab}e^ae^b$. In the case of an
integrable almost complex structure $J$ on $X^{2n}$ these equations were introduced by
Donaldson and Uhlenbeck \& Yau \cite{DDUY}.

We notice that the canonical connection $\Gamma$ on the tangent bundle of $S^6=G_2/$SU(3)
satisfies the DUY equations~(\ref{2.11}). In other words, its curvature obeys $R^{2,0}=0=R^{0,2}$
and $\om\lrcorner\,R=0$ with $\om$ given in~(\ref{2.3}).
This is easily seen from the explicit form~(\ref{2.9}) of the curvature~$R$.

\section{Twistor spaces of the six-sphere}

\noindent {\bf Twistor spaces of $S^6$. \ }
We mentioned in the introduction that one
can associate with $S^6$ two different twistor spaces, both with an integrable almost complex
structure. The larger one, Tw$(S^6)=\ $Spin(7)/U(3), belongs to the round sphere
$S^6=\ $Spin(7)/SU(4) having the full Lorentz symmetry SO(6)$\ \cong\ $SU(4)/$\Z_2$ on
the tangent spaces and the Levi-Civita connection.
The smaller twistor space, $\Zcal = G_2/$U(2), is associated with the nearly K\"ahler
coset space $S^6=G_2/$SU(3) having the canonical connection $\Gamma$ with a torsion
given in~(\ref{2.8}). The space $\Zcal$ is a complex submanifold of~Tw$(S^6)$.
Note that the Hermitian Yang-Mills equations~(\ref{2.11}) on~$S^6$ are SU(3) invariant
but not invariant under the full orthogonal group~SO(6).
This shows that the reduced twistor space~$\Zcal$ is more suitable than Tw$(S^6)$ for
a description of instantons on~$S^6$.

\bigskip

\noindent {\bf Coset representation of $\C P^2$. \ }
Let us consider the projection
\begin{equation}\label{3.1}
\pi:\quad \Zcal\ \longrightarrow\ S^6\=G_2/\mbox{SU(3)}
\end{equation}
with fibres
\begin{equation}\label{3.2}
\C P^2 = \mbox{SU(3)}/\mbox{U(2)}\ .
\end{equation}
Let $J_{\C P^2}^{}$ be a complex structure on $\C P^2$, $\{y^{\a}\}$  homogeneous
coordinates on $\C P^2$ and
\begin{equation}\label{3.3}
 \lambda^1=\frac{y^1}{y^3}\and  \lambda^2=\frac{y^2}{y^3}
\end{equation}
be local complex coordinates on the patch $\U_3=\{y^3\ne 0\}\subset\C P^2$.

One can choose as a coset representation of $\C P^2$ the matrix
\begin{equation}\label{3.4}
V\=\frac{1}{\g}\,\begin{pmatrix}W&\Lambda\\-\Lambda^\+&1\end{pmatrix}\ :=\
\frac{1}{\g}\,\begin{pmatrix}W_{11}&W_{12}&\la^1\\
W_{21}&W_{22}&\la^2\\-\bar\la^{\1}&-\bar\la^{\2}&1\end{pmatrix}
\ \in\mbox{SU(3)}\ ,
\end{equation}
where
\begin{equation}\label{3.5}
\g^2:=1+\Lambda^\+\Lambda = 1+\la^1\bar\la^{\1}+\la^2\bar\la^{\2} \und
W=W^\+=\g\cdot{\unity}_2 -\frac{1}{\g +1}\,\Lambda\Lambda^\+\ .
\end{equation}
It is a local section of the bundle SU(3)\ $\to \C P^2=\ $SU(3)$/$U(2).
{}From (\ref{3.4}) and (\ref{3.5}) it is easy to see that
\begin{equation}\label{3.6}
W\Lambda =\Lambda\and W^2=\g^2-\Lambda\Lambda^\+\qquad\Leftrightarrow\qquad
V^\+V={\unity}_3=VV^\+\ .
\end{equation}

\bigskip

\noindent{\bf Flat connection on $\Zcal$. \ }
Using the group element (\ref{3.4}) to
parametrize the typical $\C P^2$-fibre in~(\ref{3.1}), we introduce
a flat connection $\hat\Acal_0$ on the trivial bundle $G_2\times\Zcal\to\Zcal$ as
\begin{equation}\label{3.7}
\hat\Acal_0\=\hat V^\+\Acal_0\hat V + \hat V^\+\diff\hat V \quad\with \hat V=\begin{pmatrix}\bar
V&0&0\\0&1&0\\0&0&V\end{pmatrix}\in G_2\ , \quad V\in\mbox{SU(3)}\ ,
\end{equation}
where $\Acal_0$ is given in (\ref{2.6}). One gets
\begin{equation}\label{3.8}
\hat\Acal_0\=\begin{pmatrix}  \bar{\hat\Gamma}&-\hat\theta^\+&\hat B\\[2pt]
\hat\theta&0&\bar{\hat\theta}\\[2pt]
\bar{\hat{B}}&-\bar{\hat\theta}^\+&\hat\Gamma\end{pmatrix}\=
\begin{pmatrix} \bar V^\+\bar{\Gamma}\bar V{+}\bar V^\+\diff\bar V&-\bar V^\+\theta^\+&\bar V^\+ BV\\[2pt]
\theta\bar V&0&\bar{\theta}V\\[2pt]
V^\+\bar{{B}}\bar V&-V^\+\bar{\theta}^\+&V^\+\Gamma V{+}V^\+\diff V\end{pmatrix}\ ,
\end{equation}
and for the curvature $\hat\Fcal_0=\diff\hat\Acal_0 + \hat\Acal_0\wedge\hat\Acal_0$ we obtain
\begin{equation}\label{3.9}
\hat\Fcal_0\=\begin{pmatrix}\bar{\hat{R}}-{\hat\theta}^\+{\wedge}\hat\theta + \hat
B{\wedge}\bar{\hat{B}}&-(\diff\hat\theta^\+ + \bar{\hat\Gamma}{\wedge}\hat\theta^\+ +
\hat B{\wedge}\bar{\hat\theta}^\+)&\diff\hat B{+}\bar{\hat\Gamma}{\wedge}\hat B{+}\hat B{\wedge}\hat\Gamma{-}
\hat\theta^\+{\wedge}\bar{\hat\theta}\\[2pt]
\diff\hat\theta + \hat\theta{\wedge}\bar{\hat\Gamma}+\bar{\hat\theta}{\wedge}\bar{\hat
B}& -(\hat\theta{\wedge}{\hat\theta}^\+ +\bar{\hat\theta}{\wedge}\bar{\hat\theta}^\+)&
\diff\bar{\hat\theta}+\bar{\hat\theta}{\wedge}\hat\Gamma +{\hat\theta}{\wedge}\hat B \\[2pt]
\diff\bar{\hat B}{+}{\hat\Gamma}{\wedge}\bar{\hat B}{+}\bar{\hat B}{\wedge}\bar{\hat\Gamma}{-}
\bar{\hat\theta}^\+{\wedge}{\hat\theta}&-(\diff\bar{\hat\theta}^\+ + \hat\Gamma{\wedge}\bar{\hat\theta}^\+ +
\bar{\hat B}{\wedge}\hat\theta^\+ )&{\hat{R}}-\bar{\hat\theta}^\+{\wedge}\bar{\hat\theta} +
\bar{\hat B}{\wedge}\hat{{B}}
\end{pmatrix}\ .
\end{equation}

\bigskip

\noindent{\bf Maurer-Cartan equations on $\Zcal$. \ }
{}From the flatness $\hat\Fcal_0=0$ with (\ref{3.8}) and (\ref{3.9}) it follows that
\begin{equation}\label{3.10}
\bar{\hat\theta}^\+=V^\+\bar{\theta}^\+\ \Rightarrow\
\begin{pmatrix}\hat\theta^1\\\hat\theta^2\\\hat\theta^3\end{pmatrix}=\frac{1}{\g}\!
\begin{pmatrix}W_{11}&W_{12}&{-}\la^1\\
W_{21}&W_{22}&{-}\la^2\\\bar\la^{\1}&\bar\la^{\2}&1\end{pmatrix}\!
\begin{pmatrix}\theta^1\\ \theta^2\\ \theta^3\end{pmatrix} \and
\hat B=\bar V^\+\!BV=\sigma\!\begin{pmatrix}
\ 0&\sp\hat\theta^3&-\hat\theta^2\\
-\hat\theta^3&\ 0&\sp\hat\theta^1\\
\sp\hat\theta^2&-\hat\theta^1&\ 0\end{pmatrix}\ ,
\end{equation}
where $\hat\theta^{\a}$ are (1,0)-forms w.r.t.\ $\pi^*J\oplus J_{\C P^2}^{}$.
The latter is not integrable since
\begin{equation}\label{3.11}
\diff\begin{pmatrix}\hat\theta^1\\\hat\theta^2\\\hat\theta^3\end{pmatrix}+\hat\Gamma\wedge\begin{pmatrix}
\hat\theta^1\\\hat\theta^2\\\hat\theta^3\end{pmatrix}\=
\frac{2}{\sqrt{3}}\begin{pmatrix}\hat\theta^{\2}\wedge\hat\theta^{\3}\\\hat\theta^{\3}\wedge\hat\theta^{\1}\\
\hat\theta^{\1}\wedge\hat\theta^{\2}\end{pmatrix}
\qquad\Leftrightarrow\qquad \diff\hat\theta^\a +\hat\Gamma^\a_\b\wedge\hat\theta^\b\=\hat T^\a\ ,
\end{equation}
and we see a non-vanishing torsion $\hat T^\a$ with (0,2)-components. Here,
the connection on the tangent bundle $T\Zcal$ reads
\begin{equation}\label{3.12}
\hat\Gamma \= V^\+\Gamma V + V^\+\diff V\=\begin{pmatrix}
C_{11}{+}b&C_{12}&\hat\theta^4\\[2pt]
C_{21}&C_{22}{+}b&\hat\theta^5\\[2pt]
{-}\hat\theta^{\bar 4}&{-}\hat\theta^{\bar 5}&-2b\end{pmatrix}\ ,
\end{equation}
where
\begin{equation}\label{3.13}
\begin{pmatrix}C&0\\0&0\end{pmatrix} + \begin{pmatrix}b\cdot {\bf
1}_2&0\\0&-2b\end{pmatrix}\=\begin{pmatrix}
C_{11}{+}b&C_{12}&0\\
C_{21}&C_{22}{+}b&0\\
0&0&-2b\end{pmatrix}
\end{equation}
is the canonical $u(2)$-valued connection on the principal bundle $G_2\to G_2/\mbox{U(2)}=\Zcal$,
and $\hat\theta^4, \hat\theta^5$ are (1,0)-forms on the $\C P^2$ fibres of the twistor bundle~(\ref{3.1}).

\bigskip

\noindent{\bf Curvature of the connection $\hat\Gamma$. \ }
Consider the curvature $\hat R=\diff\hat\Gamma + \hat\Gamma\wedge\hat\Gamma$
of the connection $\hat\Gamma$ on $T\Zcal$ given by (\ref{3.12}) and~(\ref{3.4}).
One can easily calculate $\hat R$ from (\ref{3.9}) and obtain
\begin{align}
\hat R &\= {\small \begin{pmatrix}
F_{11}^C+\diff b-\hat\theta^{4\bar 4} & F^C_{12}-\hat\theta^{4\bar 5} &
\diff\hat\theta^{4}{+}(C_{11}{+}3b){\wedge}\hat\theta^{4}{+}C_{12}{\wedge}\hat\theta^{5}\\[2pt]
F_{21}^C-\hat\theta^{5\bar 4} & F^C_{22}+\diff b-\hat\theta^{5\bar 5} &
\diff\hat\theta^{5}{+}C_{21}{\wedge}\hat\theta^{4}{+}(C_{22}{+}3b){\wedge}\hat\theta^{5}\\[2pt]
{-}(\diff\hat\theta^{\bar 4}{+}\hat\theta^{\bar 4}{\wedge}(C_{11}{+}3b){+}\hat\theta^{\bar 5}{\wedge}C_{21}) &
{-}(\diff\hat\theta^{\bar 5}{+}\hat\theta^{\4}{\wedge}C_{12}{+}\hat\theta^{\bar 5}{\wedge}(C_{22}{+}3b)) &
-2\diff b+\hat\theta^{4\bar 4}+\hat\theta^{5\bar 5}\end{pmatrix}} \nonumber \\[8pt] \label{3.14}
&\= {\small \begin{pmatrix}
\frac{1}{3}(2\hat\theta^{1\1}-\hat\theta^{2\2}-\hat\theta^{3\3}) & \hat\theta^{1\2} & \hat\theta^{1\3} \\[2pt]
\hat\theta^{2\1} & \frac{1}{3}(-\hat\theta^{1\1}+2\hat\theta^{2\2}-\hat\theta^{3\3})& \hat\theta^{2\3} \\[2pt]
\hat\theta^{3\1} & \hat\theta^{3\2} & \frac{1}{3}(-\hat\theta^{1\1}-\hat\theta^{2\2}+2\hat\theta^{3\3})
\end{pmatrix}}\ ,
\end{align}
where
\begin{equation}\label{3.15}
F^C\=\diff C + C\wedge
C\=\begin{pmatrix}F^C_{11}&F^C_{12}\\[2pt]F^C_{21}&F^C_{22}\end{pmatrix}\ \in su(2)\ .
\end{equation}
The components of $F^C$ and $\diff b$ can be read off from~(\ref{3.14}).
Equation~(\ref{3.14}) also tells us that
\begin{equation}\label{3.16}
\diff\begin{pmatrix}\hat\theta^4\\\hat\theta^5\end{pmatrix}
+ (C+3b\cdot{\unity}_2)\wedge \begin{pmatrix}\hat\theta^4\\\hat\theta^5\end{pmatrix}\=
\begin{pmatrix}\hat\theta^{1\3}\\\hat\theta^{2\3}\end{pmatrix}\ .
\end{equation}
Together with (\ref{3.11}), this can be considered as the Maurer-Cartan equations
on $\Zcal$ for the forms $\hat\theta^A$, $A=1,\ldots,5$. Those are (1,0)-forms w.r.t.\ an
almost complex structure $\Jcal_-=\pi^*J\oplus J_{\C P^2}^{}$ on $\Zcal$ defined via
\begin{equation}\label{3.17}
\Jcal_-\hat\theta^A \= \im\,\hat\theta^A\ .
\end{equation}
The non-vanishing (0,2)-type components of the torsion $\hat T^A$ obstruct the
integrability of $\Jcal_-$.

\bigskip

\noindent{\bf Integrable almost complex structure on $\Zcal$. \ }
We may introduce a different almost complex structure $\Jcal_+$ on $\Zcal$ with the property
\begin{equation}\label{3.18}
\Jcal_+\vt^A = \im\,\vt^A \quad\for
\vt^1:=\hat\theta^1,\quad
\vt^2:=\hat\theta^2,\quad
\vt^3:=\hat\theta^{\3},\quad
\vt^4:=\hat\theta^4\and \vt^5:=\hat\theta^5
\end{equation}
and denote $\overline{\vt^A}=:\vt^{\bar A}$. Then from (\ref{3.11}) and (\ref{3.16}) we obtain
$$
\diff\vt^A + \widetilde\Gamma^A_B\wedge\vt^B \= \widetilde T^A \ ,
$$
where the connection $\widetilde\Gamma=(\widetilde\Gamma^A_B)$
and the torsion $\widetilde T{=}(\widetilde T^A)$ are given by
\begin{equation}\label{3.19}
\widetilde\Gamma\=\begin{pmatrix}
C_{11}{+}b&C_{12}&0&0&0\\C_{21}&C_{22}{+}b&0&0&0\\0&0&2b&0&0\\
0&0&0&C_{11}{+}3b&C_{12}\\0&0&0&C_{21}&C_{22}{+}3b\end{pmatrix} \and
\widetilde T\=\begin{pmatrix}\frac{2}{\sqrt{3}}\vt^{3\2}{-}\vt^{4\3}\\
{-}\frac{2}{\sqrt{3}}\vt^{3\1}{-}\vt^{5\3}\\{-}\frac{2}{\sqrt{3}}\vt^{12}{+}\vt^{4\1}{+}\vt^{5\2}\\
\vt^{13}\\ \vt^{23}\end{pmatrix}\ .
\end{equation}
Note that $\widetilde\Gamma$ is the {\it canonical\/} $u(2)$-valued connection on the tangent bundle
$T\Zcal$, and $\widetilde T$ is the torsion of $\widetilde\Gamma$. The torsion $\widetilde T^A$ in~(\ref{3.19})
has no (0,2)-components w.r.t.\ the almost complex structure $\Jcal_+$.
Therefore, $\Jcal_+$ is integrable, i.e.~$(\Zcal,\Jcal_+)$ is a complex manifold.

\section{Twistor description of instanton bundles over $S^6$}

\noindent{\bf Pulled-back curvature. \ }
Consider a complex vector bundle $E$ over $S^6$
with a connection one-form $\Acal$ having curvature $\Fcal$. Recall that $(E,\Acal)$ is
called an instanton bundle if $\Acal$ satisfies the Hermitian Yang-Mills (HYM or DUY) equations
(\ref{2.11}) which on $S^6$ can be written in the form
\begin{equation}\label{4.1}
 \Fcal^{0,2}=0\qquad\Leftrightarrow\qquad\Omega\wedge\Fcal =0\ ,
\end{equation}
\begin{equation}\label{4.2}
 \om\lrcorner\,\Fcal=0\qquad\Leftrightarrow\qquad\om\wedge\om\wedge\Fcal =0\ .
\end{equation}
Here $(\om ,\Omega )$ given in (\ref{2.3}) are forms defining on $S^6$ a nearly K\"ahler
structure. Note that, in this case, (\ref{4.2}) follows from~(\ref{4.1}) due to~(\ref{2.10}).

Consider the twistor fibration (\ref{3.1}). Let $(\widetilde E,\widetilde\Acal)=(\pi^*E,
\pi^*\Acal)$ be the pulled-back instanton bundle over $\Zcal$ with curvature
$\widetilde\Fcal =\diff\widetilde\Acal +\widetilde\Acal\wedge\widetilde\Acal$. We have
\begin{equation}\label{4.3}
\widetilde\Fcal \=
\sfrac12\,\widetilde\Fcal_{\a\b}\,\vt^{\a}\wedge\vt^{\b}+\widetilde\Fcal_{\a\bb}\,\vt^{\a}\wedge\vt^{\bb}+
\sfrac12\,\widetilde\Fcal_{\ab\bb}\,\vt^{\ab}\wedge\vt^{\bb}\=\pi^*\Fcal\ .
\end{equation}
Using the relation (\ref{3.10}) between $\theta^\a$ and $\hat\theta^\a$ as well as
the definition (\ref{3.18}) of $\vt^A$, we obtain
\begin{equation}\label{4.4}
\widetilde\Fcal_{\1\2}\=\sfrac{1}{\g}\, \bigl\{ \Fcal_{\1\2} + \la^1\Fcal_{\2\3}+ \la^2\Fcal_{\3\1}\bigr\}\ .
\end{equation}
Vanishing of $\widetilde\Fcal_{\1\2}$ for {\it all\/} values of $(\la^1, \la^2)\in \C P^2$
is equivalent to the instanton equations (\ref{4.1}) and~(\ref{4.2}).\footnote{
In contrast, for $\widetilde\Fcal_{\1\3}$ and $\widetilde\Fcal_{\2\3}$ we obtain complicated expressions
which vanish for all $\la^1, \la^2$ only if all components of the curvature $\Fcal$ vanish.
This yields the trivial case of a flat connection on $E$.}
In homogeneous coordinates $y^\a$ on $\C P^2$, this condition can be written as
\begin{equation}\label{4.5}
\widetilde\Fcal_{\1\2}=0\qquad\Leftrightarrow\qquad  y^\a\ve_{\a\b\g}\Fcal^{\b\g}=0\ ,
\end{equation}
where the indices $\ab ,\bb ,\ldots$ are raised with the metric $\delta^{\a\bb}$.

\bigskip

\noindent{\bf Correspondence of bundles. \ } Let us denote by $L_A$ vector fields on
$\Zcal$ of type (1,0) (w.r.t.\ the complex structure $\Jcal_+$) and by $L_{\bar A}$ their
complex conjugates, $A=1,\ldots,5$. Then we can introduce a rank-2 subbundle\footnote{
Recall that $T^{0,1}\Zcal$ is an integrable subbundle of the complexification of
$T\Zcal$.} $\Tcal^{0,1}_{(2)}$ of $T^{0,1}\Zcal$ spanned by $L_{\1}$ and $L_{\2}$ as well
as a rank-4 subbundle $\Tcal^{0,1}_{(4)}$ of $T^{0,1}\Zcal$ with $\{L_{\1}, L_{\2},
L_{\4}, L_{\bar 5}\}$ as a basis. Note that $\widetilde\Fcal_{\1\2}$ is the curvature of
a partial connection $\nabla^{}_{\Tcal^{0,1}_{(2)}}$ (see \cite{Raw}) along the
distribution $\Tcal^{0,1}_{(2)}$, and we can extend it to a partial connection
$\nabla^{}_{\Tcal^{0,1}_{(4)}}$ along ${\Tcal^{0,1}_{(4)}}$ by putting
$\widetilde\Acal_{\4}=0=\widetilde\Acal_{\bar 5}$. Neither $\Tcal^{0,1}_{(2)}$ nor
${\Tcal^{0,1}_{(4)}}$ is integrable as a subbundle of $T^{0,1}\Zcal$ and, therefore, we
cannot consider $\pi^*E$ as a Cauchy-Riemann (CR) bundle. This follows from the explicit
form of the torsion (\ref{3.19}) of the U(2)-structure on $\Zcal$. It would be
interesting to repeat our analysis for the other three known homogeneous nearly K\"ahler
spaces and to check whether there exist special cases where the integrability
obstructions vanish. However, this is beyond the scope of our paper which deals with
$S^6$ only.

In summary, we have the following picture:
\begin{equation}\label{4.6}
 \begin{CD}
  \pi^*E@>>>\Zcal\\
@.@V{\pi}V{\C P^2}V\\
E@>>> S^6
 \end{CD}
\end{equation}
where $(E,\nabla )$ with $\nabla =\diff +\Acal$ is a Hermitian Yang-Mills bundle with
curvature $\Fcal=\nabla^2$ satisfying (\ref{4.1}) and~(\ref{4.2}). From the above
discussion we obtain the equivalence of two assertions:

(i) $(\pi^*E,\pi^*\nabla )$ has its curvature $\widetilde\Fcal =\pi^*\Fcal$
vanishing along the distribution ${\Tcal^{0,1}_{(4)}}\subset T^{0,1}\Zcal$.

(ii) $(E,\Acal )$ is a HYM (instanton) bundle over $S^6$.

\noindent The non-integrability of the distribution $\Tcal^{0,1}_{(4)}$ means that the
HYM equations on $S^6$ are not integrable, contrary to the anti-self-dual Yang-Mills
equations on $S^4$. Hence, constructing instanton configurations in six dimensions is a
task more complicated than one might expect.

\bigskip

\noindent{\bf Relation with instantons on $\R^7$. \ }
Note that the cone $C(S^6)$ over $S^6$ with the metric
\begin{equation}\label{4.7}
 \diff s^2_7 \=\diff r^2 + r^2\diff s^2_{S^6} \quad\with  r\in\R_+
\end{equation}
is flat space, $C(S^6)=\R^7{\setminus}\{0\}$, for a proper normalization of the $S^6$ coframe $\{e^a\}$
such that $\r =1$ in~(\ref{2.10}). Employing the forms $(\om ,\Omega )$ defining the nearly K\"ahler
structure on $S^6$, we introduce on $\R^7$ the 3-form
\begin{equation}\label{4.8}
 \psi\ :=\ r^2\om\wedge\diff r + r^3\,\mbox{Im}\Omega\ .
\end{equation}
It is not difficult to show that (up to index permutation) its only nonzero coefficients are
\begin{equation}\label{4.9}
 \psi_{\ah\bh\ch} =1\quad\for (\ah\bh\ch)= (136), (426), (145), (235), (127), (347), (567)
\end{equation}
in the basis $\{\diff x^{\ah}\}$ with coordinates $x^{\ah}$ on $\R^7$ such that
$\delta_{\ah\bh}x^{\ah}x^{\bh}=r^2$. The above 3-form $\psi$ defines a $G_2$-structure on
$\R^7$, i.e.\ it is invariant under the $G_2\subset\ $SO(7) action. Its components~(\ref{4.9})
are often called octonionic structure constants.

Consider now a complex vector bundle $\Ecal$ over $\R^7$ with a connection $\Acal^\prime$
and curvature $\Fcal^\prime$. Employing the Hodge operator~$*$ in~$\R^7$,
we impose on $\Acal^\prime$ the first-order differential equations
\begin{equation}\label{4.10}
 *\psi\wedge\Fcal^\prime =0\qquad\Leftrightarrow\qquad\psi_{\ah\bh\ch}\Fcal^{\prime\,\bh\ch}=0
\end{equation}
which are called $G_2$-instanton equations~\cite{DT}. Their solutions automatically
satisfy the Yang-Mills equations on~$\R^7$.

It was shown by Tian \cite{Tian} that solutions $\Acal^\prime$ of (\ref{4.10}) obeying also
\begin{equation}\label{4.11}
\partial_r\lrcorner\,\Acal^\prime =0\und\partial_r\lrcorner\,\Fcal^\prime =0
\end{equation}
are equivalent to solutions $\Acal$ of the HYM equations (\ref{4.1}) and (\ref{4.2}) on $S^6$.
He calls such configurations {\it tangent instantons\/} on~$\R^7$. Examples of such instanton solutions were
discussed in~\cite{HILP, HN}.

A twistor description of solutions to~(\ref{4.10}) on any 7-dimensional Riemannian manifold~$X$
with $G_2$-holonomy was recently proposed by Verbitsky~\cite{Ver}. Namely, he introduced a so-called
CR~twistor space of~$X$ as the bundle $\pi : S^6X\to X$ of unit six-spheres in the tangent bundle~$TX$.
For $\R^7$ this space is a direct product manifold
\begin{equation}\label{4.12}
 {\rm Tw}(\R^7)\=\R^7\times S^6\ .
\end{equation}
It was shown \cite{Ver} that the complexified tangent bundle of Tw$(X)$ has an integrable
complex rank-3 subbundle $\Tcal^{0,1}_{(3)}$ if $X$ is a $G_2$-holonomy manifold. For a
bundle $\Ecal$ over $X$ with a connection $\Acal^\prime$ one can introduce the
pulled-back bundle $(\pi^*\Ecal , \pi^*\Acal^\prime)$ over Tw$(X)$. It was proven that
$G_2$-instanton bundles over $X$ correspond to CR-bundles over Tw$(X)$ with a flat partial
(0,1)-connection $\bar\partial_{\pi^*\Ecal}$ defined on the distribution
$\Tcal^{0,1}_{(3)}$ \cite{Ver}. In other words, the $G_2$-instanton equations on $X$ are
equivalent to the equations $\bar\partial_{\pi^*\Ecal}^2=0$ on Tw$(X)$. This theorem
obviously applies to the case of~$X=\R^7$. Specializing then to tangent solutions
(in the sense of~(\ref{4.11})) to the $G_2$-instanton equations~(\ref{4.10}) on~$\R^7$
will yield solutions of the HYM equations on~$S^6$.
Thus, the twistor description of instantons on~$S^6$ is related to
the twistor description of $G_2$-instanton solutions on~$\R^7$.

\paragraph{Acknowledgments.}
We would like to thank Vicente Cort\'es, Klaus Hulek and Misha Verbitsky for useful
discussions in the framework of the Graduiertenkolleg ``Analysis, Geometry and String Theory.''
This work was partially supported by the Deutsche Forschungsgemeinschaft
and the Riemann Center for Geometry and Physics of Leibniz Universit\"at Hannover.

\end{document}